\documentclass[%
reprint,
superscriptaddress,
twocolumn,
notitlepage,
nofootinbib,
amsmath,amssymb,amsfonts
floatfix
]{revtex4-1}
\usepackage{mathtools}
\usepackage{amsthm}
\usepackage{amssymb}
\usepackage{amsfonts}
\usepackage{amsmath}
\usepackage{appendix}
\usepackage{physics}
\usepackage{graphicx}
\usepackage{dcolumn}
\usepackage{bm}
\usepackage{hyperref}
\usepackage[normalem]{ulem}
\usepackage{xcolor}
\usepackage{times}
\usepackage[utf8]{inputenc}
\usepackage{graphicx}
\usepackage{float}
\usepackage{caption}
\usepackage{subcaption}

\newtheorem{rem}{Remark}

\pdfoutput=1

\begin{document}


\title{Quantum state features of the FEL radiation from the occupation number statistics}

\author{F. Benatti}
\affiliation{Dipartimento di Fisica, Universit\`a degli Studi  di Trieste, I-34151 Trieste, Italy}
\email{corresponding authors: Fabio Benatti (benatti@ts.infn.it) theory and Fulvio Parmigiani (fulvio.parmigiani@elettra.eu) experiment}
\affiliation{INFN, Sezione di Trieste, I-34151 Trieste, Italy}
\author{S. Olivares}
\affiliation{Dipartimento di Fisica “Aldo Pontremoli,” Università degli Studi di Milano, I-20133 Milan, Italy}
\affiliation{INFN, Sezione di Milano, I-20133 Milan, Italy}
\author{G. Perosa}
\affiliation{Dipartimento di Fisica, Universit\`a degli Studi  di Trieste, I-34151 Trieste, Italy}
\affiliation{Elettra Sincrotrone Trieste, I-34149 Basovizza, Trieste, Italy}
\author{D. Bajoni}
\affiliation{Dipartimento di Ingegneria Industriale e dell’Informazione,
Universit\`a degli Studi di Pavia, I-27100 Pavia, Italy}
\author{S. Di Mitri}
\affiliation{Dipartimento di Fisica, Universit\`a degli Studi  di Trieste, I-34151 Trieste, Italy}
\affiliation{Elettra Sincrotrone Trieste, I-34149 Basovizza, Trieste, Italy}
\author{R. Floreanini}
\affiliation{INFN, Sezione di Trieste, I-34151 Trieste, Italy}
\author{L. Ratti}
\affiliation{Dipartimento di Ingegneria Industriale e dell’Informazione,
Universit\`a degli Studi di Pavia, I-27100 Pavia, Italy}
\affiliation{INFN Sezione di Pavia, I-27100 Pavia, Italy}
\author{F. Parmigiani}
\affiliation{Dipartimento di Fisica, Universit\`a degli Studi  di Trieste, I-34151 Trieste, Italy}
\affiliation{Elettra Sincrotrone Trieste, I-34149 Basovizza, Trieste, Italy}
\affiliation{International Faculty, University of Cologne, Albertus-Magnus-Platz, 50923 Cologne, Germany}

\begin{abstract}
The statistical features of the radiation emitted by Free-Electron Lasers (FELs), either by Self-Amplified Spontaneous Emission (SASE-FELs) or by seeded emission (seeded-FELs), are attracting increasing attention because of the use of such light in probing high energy states of matter and their dynamics.
While the experimental studies conducted so far have mainly concentrated on correlation functions, here we shift the focus towards reconstructing the distribution of the occupation numbers of the radiation energy states. In order to avoid the various drawbacks related to photon counting techniques when large numbers of photons are involved, we propose  a Maximum Likelihood reconstruction of the diagonal elements of the FEL radiation states in the energy eigenbasis based on the statistics of no-click events.
The ultimate purpose of such a novel approach to FEL radiation statistics is the experimental confirmation that SASE-FEL radiation exhibits thermal occupation number statistics, while seeded-FEL light Poissonian statistics typical of coherent states and thus of laser light. In this framework, it is interesting to note that the outcome of this work can be extended to any process of harmonic generation from a coherent light pulse, unlocking the gate to the study of the degree to which the original distinctive quantum features deduced from the statistical photon number fluctuations are preserved in non-linear optical processes.
\end{abstract}

\maketitle



\section{Introduction}

\begin{figure*}[t]
\centering
\includegraphics[width=17.8cm]{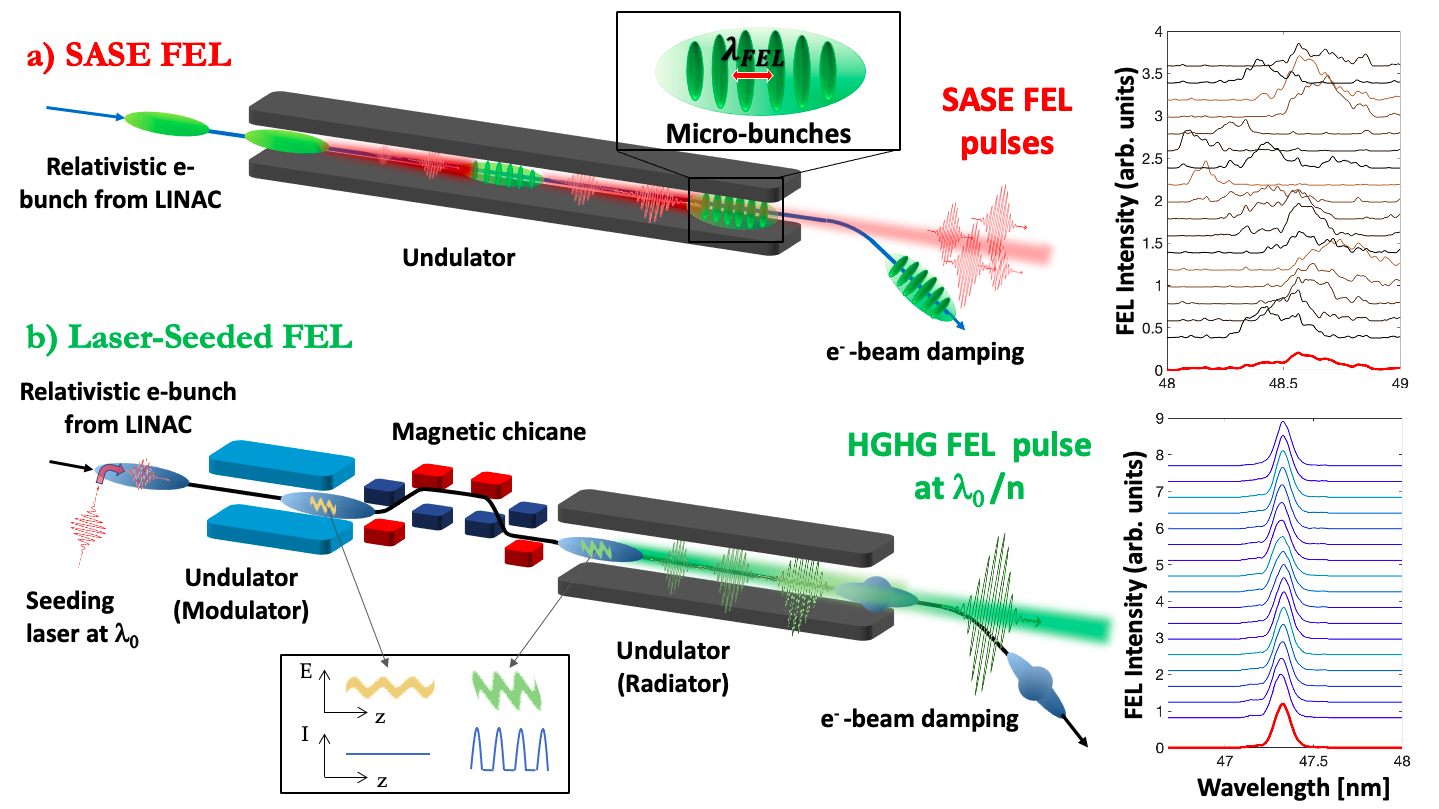}
\captionsetup{justification=raggedright}
\caption{Left: cartoon of lasing in SASE and HGHG mode. In SASE, energy and density modulation of electrons at the scale of the resonant FEL wavelength happen simultaneously along the undulator. The spectrum of the amplified spontaneous radiation is an ensemble of spikes whose intensity varies shot to shot, each spike being individually coherent. In HGHG, the energy modulation generated by the electron-laser interaction in the modulator translates into density modulation via the magnetic chicane. The FEL spectrum is a narrow band single spike. Right: intensity profile of consecutive shots of the spectrum of the FERMI FEL, run in SASE (top) and HGHG mode (bottom), for the same setting of the accelerator.}
\label{FEL}
\end{figure*}

Recent years have seen an increasing interest in  the generation of ultra-bright and ultrashort electromagnetic pulses by means of radiation sources, widely known as Free-Electron Lasers (FELs), through self-amplification of spontaneous emission (SASE) from a relativistic electron  bunch wiggling between a series of magnetic poles known as undulators~\cite{Schmueser}. 
Despite the growing experimental and theoretical activity in the field, the issue concerning the nature of statistical fluctuations in the emitted radiation remains still largely unmet.
In the extreme, such a complex issue boils down to the question: "To what degree is FEL radiation laser radiation?”

The experimental intensity, angular and spectral distributions of the radiation emitted by SASE-FELs are well predicted by classical electrodynamics. Not surprisingly, also a quantum field theory approach, such as that firstly proposed in~\cite{Preparata}, can recover the well-known exponential growth of the electromagnetic  field intensity along with the exponential regime yielding a satisfactory description of the behavior of a single passage SASE-FEL. 
Instead, more surprising theoretical predictions appear in~\cite{Becker} and~\cite{Gjaja} that the radiation resulting from a SASE-FEL could show properties typical of squeezed light. 
Indeed, any quantum description of an electromagnetic phenomenon also contains a classical description as an approximation.
On the contrary, a phenomenon which can be described by a classical theory, and yet exhibits purely quantum features, points to a weakness in the classical explanation and requires a quantum theory of SASE-FEL radiation that is able to reproduce not only its classical features, but also the quantum characteristics. Although a possible explanation might be found in the fact that the spontaneous emission of radiation originates from the fluctuations of the electromagnetic vacuum,  
a solid theoretical explanation of such an effect is, to the best of our knowledge,  still missing. 

In this respect,  the study provided in~\cite{Chen} made perfect sense; there, in a photon counting experiment performed on the spontaneous harmonic radiation generated by an infrared, SASE-FEL using a well-defined ensemble of electron pulses, the authors announced the detection of genuinely quantum sub-Poissonian intensity fluctuations.
However, a more recent work~\cite{Park} challenges such findings together with their consequences in that the reported quantum signatures of the SASE-FEL radiation can be explained, within standard, classical SASE-FEL theory, by combining detector dead-time effects with photon clustering arising from the FEL gain. 
Thus, to bridge the still existing gap between the theoretical predictions of quantum effects and the experimental evidence, one must develop further a fully consistent quantum explanation of SASE-FEL radiation production that predicts possible distinctive quantum features in the statistical fluctuations of SASE-FEL radiation.
 
A completely different scenario arises if the FEL process is triggered not by the background radiation of an oscillating relativistic electron bunch, but by a pulse of laser light with intensity several orders of magnitude larger than the background incoherent shot noise. At the moment, we can only conjecture that the coherent features of the "seeding" laser pulse are transferred to the electrons in the bunch that in turn convey them, at least in part, to the FEL radiation, first in the modulation phase and then in the radiation phase, thus producing amplified high harmonic radiation extending up to the X-ray domain and showing a coherent, that is a lasing state, of the electromagnetic field.

However, to be coherent and thus to describe laser light, the quantum states of seeded-FEL radiation must satisfy with specific requisites: they must be characterized by all $2n$-point normalized Glauber correlation functions $g_n$ equal to $1$ in modulus~\cite{Glauber} and the occupation numbers 
of the available energy states must exhibit a Poisson distribution.
Experimental investigations focussing upon the Glauber correlation functions $g_1$ and $g_2$ both for SASE-FEL~\cite{Singer,Roling} and seeded-FEL light~\cite{Elettra1}
revealed typical chaotic light behaviour in the first case, $\vert g_2\vert =1+\vert g_1\vert^2$, and  values of $\vert g_2\vert$ being still greater than, but rather close to $1$ in the second one. However, $\vert g_1\vert=\vert g_2\vert=1$ would only confirm the possibility of lasing features, but does not guarantee them. For such a purpose, higher order 
$g_n$ should be investigated. In the following, we instead propose to shift the focus from Glauber correlation functions to reconstructing the statistical properties of the FEL radiation related to the photon occupation numbers.

\section{Concepts and Methods}

The quantum nature of the FEL radiation can be proved either with specific experiments aimed at highlighting indirect clues of its granularity~\cite{Eberly} or by retrieving the actual photon-number distribution of the field \cite{Mandel}. In the latter case, photon-counting techniques have many drawbacks, as evidenced by~\cite{Park} in the case of~\cite{Chen}. In general, they are limited to the detection of up to few tens of photons~\cite{Allevi:hyb,chesi,Perina12,Dovrat13,Avenhaus10}. If the average number of photons increases, as in the case of high intensity sources, photon counting becomes challenging, even for attenuated FEL radiation.  For the reconstruction purposes outlined above, one has then to resort to other approaches, such as Maximum Likelihood (ML)~\cite{Dempster,Vardi,Boyles} methods based on no-click probabilities instead of photon counting. Finally, within the present context, it is interesting to note that the question concerning the degree to which the statistical occupation of the modes of a laser pulse is preserved after a nonlinear conversion of the photon energy remains an unknown aspect that could find an adequate answer via the theory and detection technique reported here.

In the following, we pursue the study of the photonic structure of the light emitted by FELs using the statistics of no-click events in Geiger-like photo detectors~\cite{Mogilevtsev, Brida11, Rossi}. Such an approach has already been successfully investigated and experimentally tested in the optical regime, both for continuous-wave and pulsed sources~\cite{Zambra,Allevi,Brida2}, but still limited to  a mean number of only tens of photons. Therefore, to apply the technique successfully to (attenuated) FEL light, in the following section we show by simulated experiments that it can be extended to reconstruct the statistics of hundreds of photons. 
Notice that such a procedure provides information about the photon number statistics of the state of the light generated by the FEL, without any assumption about the nature of the FEL process itself. 
From this point of view, one expects a nearly Poissonian distribution of the occupation number  in the case of seeded FEL light and  a substantially thermal one in the SASE case. 
Shifting the focus from photon correlations to the statistics of occupation numbers would then provide a first step towards a reliable investigation of the quantum signatures of FEL light, ultimately able to discriminate between Poissonian and non-Poissonian features.

Notice that the experimental evidence of classical, non-Poissonian statistics, such as a thermal distribution,  would exclude lasing properties.
On the other hand, the experimental confirmation of the presence of Poissonian statistics, though not sufficient, would in any case provide solid ground to attributing "lasing" to the mechanism behind the generation of radiation by relativistic electrons wiggling in a series of magnetic devices.
It would also give support to the conjecture that the lasing properties are imprinted upon the seeded-FEL radiation by the coherence features of the seeding pulse. 
Furthermore, an efficient applicability of the reconstruction mechanism to the diagonal occupation number statistics of the FEL radiation would boost further steps towards 
the full reconstruction~\cite{Brida11} of the quantum states of the FEL radiation and prompt the investigation of which quantum features the emitted radiation would exhibit, 
should one use squeezed, namely genuinely quantum seeding pulses.  

From a more general perspective, the consequences of the confirmation of the lasing properties of the seeded-FEL light will be of two types: it will open the way to experiments in quantum optics with X-ray radiation and it will foster the experimental investigation  of light-matter interactions in the high energy quantum regime.

\subsection*{FEL radiation: quantum states}    

We start with an overview of the quantum states of light that are most suited to describe SASE and seeded-FEL radiation.

All light is in principle describable by non-commuting creation and annihilation operators $a$ and $a^\dag$ of electromagnetic energy modes of frequency $\omega$ such that $[a\,,\,a^\dag]=1$. The number operator is $N=a^\dag a$ and the so-called Fock number states with $n$ photons,  
$\vert n\rangle = (n!)^{-\frac12} (a^\dag)^n\vert 0\rangle$, satisfy $N\vert n\rangle=n\vert n\rangle$ and constitute an orthonormal basis in the Fock Hilbert space 
associated with the chosen mode.                    

Let $\langle\,X\,\rangle={\rm Tr}\Big(\rho\,X\Big)$ denote the expectation value of an electromagnetic field operator $X$ with respect to the field state (density matrix) $\rho$, which, in the number state representation, is in general a non-diagonal positive matrix of trace $1$,
\begin{equation}
\label{rhonondiag}
\rho=\sum_{k,\ell=0}^\infty \rho_{k\ell}\,\vert k\rangle\langle \ell\vert\ , \quad \sum_{\ell=0}^\infty\rho_{\ell\ell}=1\ ,
\end{equation}
where $\rho_{k\ell}=\langle k\vert\rho\vert\ell\rangle$.
Let us consider a monochromatic  field polarized along the $z$ axis and propagating along the $x$ direction, with energy $\omega$ ($\hbar=1$), wave-vector $k=\omega/c$, and right, respectively, left moving field components
\begin{equation}
\label{lrmcomp}
E_+(x,t)=i\,\frac{\omega}{c}\,{\rm e}^{i(kx-\omega t)}\,a\ ,\quad E_-(x,t)=E^\dag_+(x,t)\ .
\end{equation}

At a fixed spatial position $x$, the Glauber correlation functions of order $n$~\cite{Glauber}, depend on $n$ varying instants of times and are given by
\begin{equation}
\label{Gnn}
g_n\left(\left\{t_i\right\}_{i=1}^n\right)\equiv\frac{G_n\left(\left\{t_i\right\}_{i=1}^n\right)}{\prod_{j_=1}^{2n}\sqrt{G_1(t_j)}}\ ,
\end{equation}
with
\begin{equation}
\label{Gn}
G_n\left(\left\{t_i\right\}_{i=1}^n\right)=\left\langle\,E_-(t_1)\cdots E_-(t_n)
E_+(t_{n+1})\cdots E_+(t_{2n})\,\right\rangle,
\end{equation} 
where we have disregarded the spatial dependence.
By suitably choosing the times $t_i$, one also eliminates the time-dependence and obtains~\cite{Mandel}
\begin{equation}
\label{g1TD}
g_1(0)=1\ ,\quad 
g_2(0)=1+\frac{\langle N^2\rangle-\langle N\rangle^2-\langle N\rangle}{\langle N\rangle^2}\ .
\end{equation}                   

Monochromatic coherent states $\vert\alpha\rangle$ with complex amplitudes $\alpha\in\mathbb{C}$ such that $a\vert\alpha\rangle=\alpha\vert\alpha\rangle$, exhibit Poissonian occupation number distributions. They are generated by displacing the vacuum state $\vert\alpha\rangle=D(\alpha)\vert0\rangle$ by means of the displacement operator $D(\alpha)=\exp(\alpha a^\dag-\alpha^* a)$ (see the Supplementary Material for some of their properties). Coherent states yield mean-values $\langle\alpha\vert (a^\dag)^n a^n\vert\alpha\rangle=|\alpha|^{2n}$ whence $\vert g_n(0)\vert=1$ for all $n\geq 1$. The same do the
uniform averages of projectors onto coherent states over $\varphi$ in $\displaystyle \alpha=|\alpha|{\rm e}^{i\varphi}$ , which are commonly used to describe the quantum states of laser light for which the phases are not accessible. 

Instead, thermal states at temperature $T$ setting the Boltzamnn constant $\kappa_B=1$),
\begin{equation}
\label{Gibbs}
\hskip -.3cm
\rho_{T}=\sum_{k=0}^\infty\frac{n^k_T}{(1+n_{T})^{k+1}}\vert k\rangle\langle k\vert,\ n_T\equiv\langle N\rangle=\frac{1}{{\rm e}^{\omega/T}-1}\ ,
\end{equation}
give $\langle N^2\rangle-\langle N\rangle=2\,\langle N\rangle^2$ and thus $g_2(0)=2$ corresponding to the incoherent character of such states of radiation.      

The randomness of thermal states  can be lessened  by displacing them:
\begin{equation}
\label{displth1}
\rho_{T,\alpha}:=D(\alpha)\,\rho_T\,D^\dag(\alpha)\ ,
\end{equation}
whence
$\langle N\rangle={\rm Tr}\left(\rho_{T,\alpha}\,a^\dag a\right)=n_T+|\alpha|^2$. Thus, 
\begin{equation}
\label{g2Talpha}
\vert g_2(0)\vert=1+\frac{n_T}{|\alpha|^2}\frac{1}{(1+\frac{n_T}{|\alpha|^2})^2}+\frac{n_T}{|\alpha|^2}\frac{1}{1+\frac{n_T}{|\alpha|^2}}\leq 2\ ,
\end{equation}
reaches the minimum value $1$ when $n_T=0$, namely for purely coherent  states and the maximum value $2$ when $\alpha=0$, that is for purely thermal states, or in the limit of infinite $n_T$.

Displaced thermal states are natural candidates to describe the radiation emitted by FELs, which is due to aggregation in micro-bunches of relativistic electrons wiggling transversally while passing through a magnetic undulator. The micro-bunching is induced by the electrons interacting  with the electric field they generate.
SASE-FELs undulator radiation is the result of interaction between the electrons and the virtual photons of the static magnetic field.  The spiky behavior is due to the lack of the causal connection between segments separated by cooperation length due to the different forward velocity between the electron bunch and the e.m. radiation field \cite{Bonifacio2}.
To lessen the initial randomness of the electron bunch, seeded-FELs instead use an external laser radiation source to "order" the electrons before the undulator, 
with the effect that the various micro-bunches emit in phase within a very narrow spectral band-width, thus introducing coherence with respect to the thermal case.

\subsection*{No-click quantum state reconstruction} 

As anticipated in the Introduction, this work proposes to shift focus from the Glauber correlation functions to the occupation number statistics. Such statistics are given by the diagonal entries  $\rho_k:=\rho_{kk}$ of the quantum state of light $\rho$ in the Fock number state representation (see~\eqref{rhonondiag}). The statistics will  be reconstructed by means of the empirically measured  probability of no-click events, thus avoiding the drawbacks of photon counting. 
The principal tool is a Maximal Likelihood algorithm, as proposed first by E. Fermi ~\cite{Fermi}, based on the \textit{no-click} probabilities
\begin{equation}
\label{NoclickProb}
P(\{\rho_n\}_n,\eta)=\sum_{k=0}^\infty(1-\eta)^k\,\rho_k\  ,
\end{equation}
where $\eta$ is the detector efficiency, namely the probability that it clicks when impinged by a single photon.
The no-click probability for $k$ impinging photons is given by $(1-\eta)^k$, whence~\eqref{NoclickProb} gives  the total no-click probability.

Since higher photon-number states have reduced populations, one truncates the sum at $N$ such that $\sum_{k=1}^N\,\rho_k\geq 1-\epsilon$ for a sufficiently small $\epsilon$.
Photon detection at $L$ different efficiencies  $\{\eta_\ell\}_{\ell=1}^L$ yields a system of $L$ linear equations 
\begin{equation}
P_\ell(\{\rho_n\}_n):=P(\{\rho_n\}_n,\eta_\ell)=\sum_{k=0}^N(1-\eta_\ell)^k\,\rho_k\ ,
\label{NoclickModel}
\end{equation}
that can be used to infer the unknown parameters $\rho_n$ via a Maximum Estimation method (see the Supplementary Material). The number of efficiencies $L$ being in general much smaller than $N$, the ensuing underestimation and the existence of more than one solution can be overcome by applying an external energy constraint 
$\beta$ (see  the Supplementary Material for a short discussion of this point). 

In the non-monochromatic case with $M>1$ different independent modes, the quantum state of the beam is of the form $\rho=\bigotimes_{j=1}^M\rho^{(j)}$, 
where $\rho^{(j)}$ denotes the state of the $j$-th mode. Then, the joint probability for $\bar n$ impinging photons and no clicks reads
\begin{equation}
\label{multimodeprob0}
\Pi_{\bar n}(M,\eta)=\sum_{n_1+n_2+\cdots+n_M=\bar n}\,\left[\prod_{j=1}^M(1-\eta)^{n_j}\rho^{(j)}_{n_j}\right]\ ,
\end{equation}
where $n_j$ is the occupation number of mode $j$ and $\rho^{(j)}_{n_j}$ is the probability of having $n_j$ photons given the state $\rho^{(j)}$. Then, the total multi-mode no-click probabilities become 
\begin{equation}
\label{multimodeprob0a}
P_M(\eta)=\sum_{\bar n=0}^\infty \Pi_{\bar n}(M,\eta)\ .
\end{equation}

\begin{figure*}[hbt!]
\centering
\begin{subfigure}[b]{0.48\textwidth}
\centering
\includegraphics[width=0.95\textwidth]{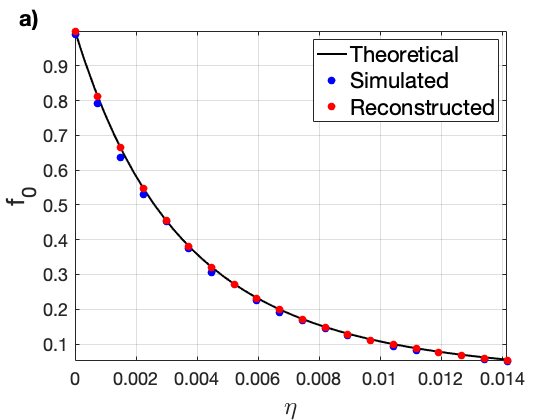}
\end{subfigure}
\begin{subfigure}[b]{0.48\textwidth}
\centering
\includegraphics[width=0.95\textwidth]{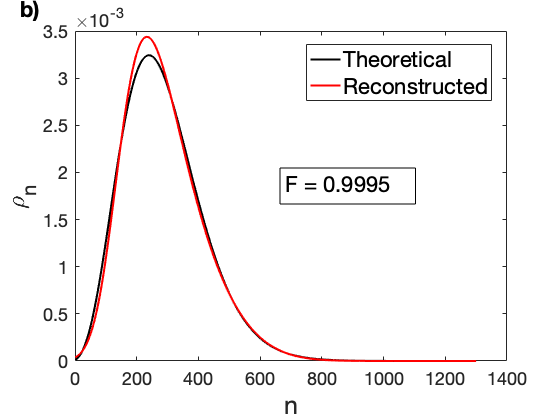}
\end{subfigure}
\captionsetup{justification=raggedright}
\caption{Single-mode displaced thermal state, $M=1$ $\alpha = 16$, $n_T = 30$. Figure a: simulated no-click probabilities (blue dots), outputs of the reconstruction algorithm (red dots), plot of~\eqref{NoclickProb} (solid line).\\
Figure b: reconstructed photon distribution (red line), photon distribution from~\eqref{NoclickProb} (black line). Fidelity $F$ also shown.}
\label{Seeded1}
\vspace{4ex}
\centering
\begin{subfigure}[b]{0.48\textwidth}
\centering
\includegraphics[width=0.95\textwidth]{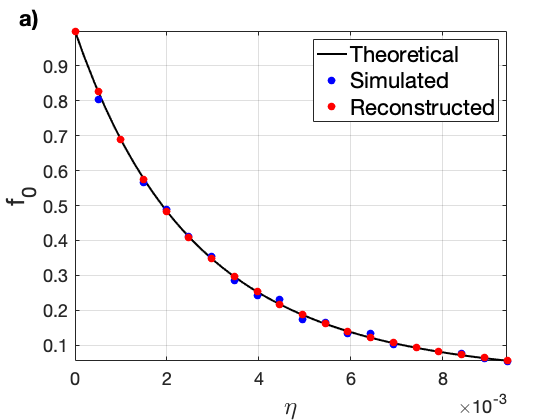}
\end{subfigure}
\begin{subfigure}[b]{0.48\textwidth}
\centering
\includegraphics[width=0.95\textwidth]{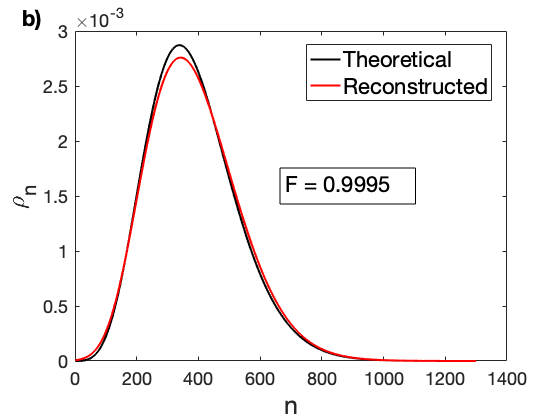}
\end{subfigure}
\captionsetup{justification=raggedright}
\caption{Multi-mode displaced thermal state: $M = 3$, $\alpha = [7,8,12]$ and $n_T = 30$. Figure a: simulated no-click probabilities (blue dots), outputs of the reconstruction algorithm (red dots), plot of~\eqref{multimodeprob0a} (solid line).\\
Figure b: reconstructed photon distribution (red line), photon distribution from~\eqref{multimodeprob0a} (black line). Fidelity $F$ also shown.}
\label{Seeded2}
\end{figure*}

					
The reconstruction algorithm described in the Supplementary Material is such that,  in the monochromatic case, 
once fed with  the no-click probabilities 
$P_\ell(\{\rho_n\}_n)$ measured for sufficiently many detector efficiencies $\eta_\ell$, $1\leq\ell\leq L$,  it returns the populations $\rho_k$ of the $k$-photon 
Fock states of light, from $k=0$ up to $k=N$ (see the right plot in Figure~\ref{Seeded1}).  Instead, in the case of multi-mode radiation, from the no-click 
probabilities $P_M(\eta)$ the algorithm is able to reconstruct the probabilities in~\eqref{multimodeprob0}, denoted by $\Pi_n$ on the $y$ axis of the right plots in 
Figures~\ref{Seeded2}-\ref{SASE2}. 
To retrieve the single-mode probabilities $\rho^{(j)}_{n_j}$ and thus the diagonal elements of the multi-mode state $\rho$, further considerations  are needed  
based on suitable physical assumptions about the structure of the radiation quantum state. 


\section{Numerical simulations} 	

To check the reconstruction algorithm, one starts from known population distributions $\{\rho^{in}_n\}_n$, them simulates 
the no-click probabilities for  chosen efficiencies, and feeds them to the algorithm thus obtaining the reconstructed $\{\rho^{out}_n\}_n$, from which  the fidelity 
\begin{equation}
\label{Fidelity}
F = \sum_{j = 1}^N \sqrt{\rho^{in}_j\rho^{out}_j }\ ,
\end{equation}
can be evaluated.
In the following we focus on the simulation of two types of radiation quantum states: a single-mode displaced-thermal state and a multi-mode  displaced-thermal state of the form 
\begin{equation}
\label{multithdisp}
\rho=\bigotimes_{j=1}^M\rho^{(j)}_{T,\alpha_j}\ .
\end{equation}
with a common temperature $T$ and different amplitudes $\alpha_j$.    

As previously discussed, the first kind of quantum state should represent well the light produced by a seeded FEL.  
For SASE-FELs instead, the common physical context at the origin  of the micro-bunches suggests describing the corresponding quantum state of light by the multi-mode state~\eqref{multithdisp} with a same degree of randomness and hence the same temperature $T$, but with different coherent biases and thus with different displacement amplitudes  
$\alpha_j$.

\begin{figure*}[t!]
\centering
\begin{subfigure}[b]{0.48\textwidth}
\centering
\includegraphics[width=0.95\textwidth]{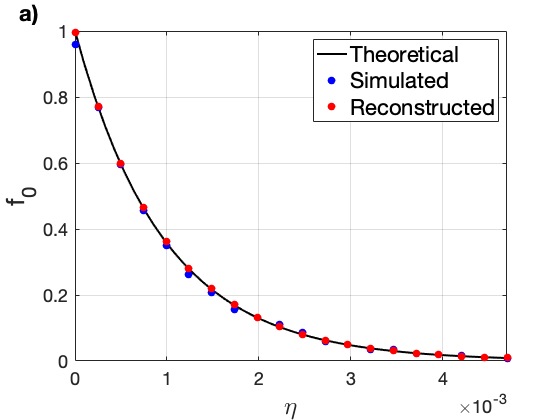}
\end{subfigure}
\begin{subfigure}[b]{0.48\textwidth}
\centering
\includegraphics[width=0.95\textwidth]{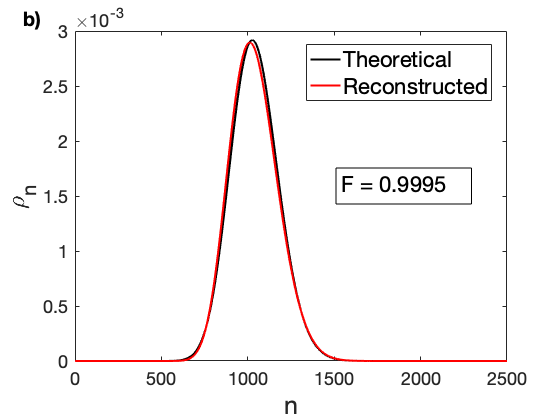}
\end{subfigure}
\captionsetup{justification=raggedright}
\caption{Multimode-displaced thermal state: $M = 30$, with $\alpha_j$ chosen randomly in the range $[1,8]$ and $n_T = 10$. Figure a: simulated no-click probabilities (blue dots), outputs of the reconstruction algorithm (red dots), plot of~\eqref{multimodeprob0a} (solid line).
Figure b: reconstructed photon distribution (red line), photon distribution from~\eqref{multimodeprob0a} (black line). Fidelity $F$ also shown.}
\label{SASE1}
\vspace{4ex}
\centering
\begin{subfigure}[b]{0.48\textwidth}
\centering
\includegraphics[width=0.95\textwidth]{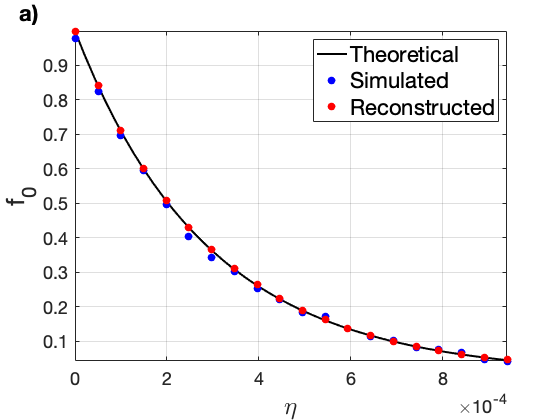}
\end{subfigure}
\begin{subfigure}[b]{0.48\textwidth}
\centering
\includegraphics[width=0.95\textwidth]{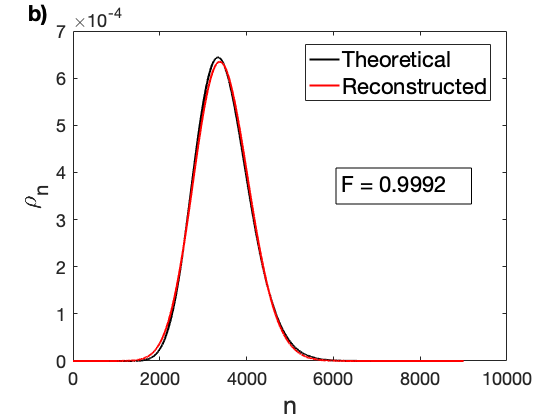}
\end{subfigure}
\captionsetup{justification=raggedright}
\caption{Multimode-displaced thermal state: $M = 30$, $\alpha_j$ chosen randomly in the range $[1,5]$ and $n_T = 100$.Figure a: simulated no-click probabilities (blue dots), outputs of the reconstruction algorithm (red dots), plot of~\eqref{multimodeprob0a} (solid line).\\
Figure b: reconstructed photon distribution (red line), photon distribution from~\eqref{multimodeprob0a} (black line). Fidelity $F$ also shown.}
\label{SASE2}
\end{figure*}
					
Due to the one-dimensional feature of the problem, the generation of a vector of no-click probabilities can be obtained by generating events according to the probabilities $P_\ell(\{ \rho_n\}_n)$ by means of the rejection sampling technique \cite{SRM}. 
Then, the occurrences of the generated points, normalized to the total number of data, coincide with the desired no-click probabilities that, for sake of simplicity, are 
denoted by $P$ on the $y$ axis of the left plots in Figures~\ref{Seeded1}-\ref{SASE2}. From datasets of thousands of values, we extract a subset of $L$ elements with the corresponding efficiencies. \\
In the following, $L$ is chosen to be equal to $20$. Such choice is consistent with previous tests of the algorithm~\cite{Genovese,Brida} and with possible experimental set-ups, as described below. The range of selected efficiencies is chosen taking particular care to avoid regions of $\eta$ where the no-click probabilities are flat. This condition leads to the exclusion of points with low Fisher information.

To mimic a realistic experiment, random perturbations have been added to the no-click probabilities. The stability of the algorithm has also been tested by corrupting the simulated no-click probabilities with systematic perturbations.
The iteration procedure described in the Supplementary Material has then been fed with the no-click probabilities and the initial occupation numbers $\rho_n(0)$ taken all equal: $\rho_n(0)=1/N$. 
Finally, the consistency of the photon distribution reconstructed by the algorithm with the theoretical one is evaluated by means of the fidelity~\eqref{Fidelity}.\\
As mentioned above, the algorithm has been applied to two distinct dataset scenarios. The dataset parameters have been chosen in adherence to the performance of a short wavelength high gain FEL operating in the SASE \cite{Bonifacio1, Kondratenko} and in an external seeding  \cite{Yu} regime, respectively. The FEL pulse photon distribution has been assumed to be taken close to power saturation in both regimes.

Extreme ultraviolet seeded FELs are ideally capable of producing Fourier-transform limited light pulses, hence a single transverse and longitudinal mode. The natural relative spectral bandwidth can be as low as $10^{-4}$ \cite{Allaria1, Allaria2}. Thus, micron-scale modulations of the electron energy  distribution, accumulated during the acceleration process, can add a statistically significant pedestal to the FEL spectrum \cite{Marcus, Hemsing, Zhang, Perosa}.

The ideal and the perturbed seeded FEL performance is represented in our algorithm through a single mode-displaced thermal state $(M=1, \alpha=16)$, and a multi-mode-displaced thermal state $(M=3, \alpha=7,8,12)$. In both cases, the mean occupation number has been chosen $n_T=30$. The reconstructed no-click frequencies and photon distributions are compared with the theoretical and simulated data in Figures \ref{Seeded1} and \ref{Seeded2}.

SASE-FELs are driven by shot noise in the electron distribution. Spontaneous undulator radiation is amplified and, eventually, a large number of longitudinal modes - each individually coherent - is emitted. A single transverse mode is instead selected by the amplification process close to saturation~\cite{Saldin}. 
The number of longitudinal modes is basically given~\cite{Bonifacio2} by the portion of bunch length contribution to FEL emission ($\sim 100$ $\mu$m) divided by the FEL coherence length  ($\sim 1-3$ $\mu$ in the extreme ultraviolet regime).

The SASE-FEL radiation has  therefore been modelled choosing $M=30$ and randomly distributed displacements. Moreover, two thermal configurations were considered to test the algorithm: in the first example the amplitudes $\alpha_j$ are chosen in the interval $1-8$ with $n_T=10$ and in the interval $1-5$ with $n_T=100$ in the second one. Results are shown in Figures \ref{SASE1} and \ref{SASE2}.
					
\indent
The overall procedure requires less than 200 iterations, to reach a fidelity level, $1-F$, smaller than $10^{-4}$. Similarly high fidelities have been obtained for a 
larger number of datasets, thus demonstrating the capability of the  algorithm to retrieve the input diagonal statistics in a wide range of configurations.
In Figures~\ref{Seeded1}--\ref{SASE2}, the left plots display a comparison between simulated, reconstructed and theoretical no-click frequencies, while the right ones compare reconstructed and theoretical photon distributions. More specifically, the left plots in each figure show the dependence  of no-click probabilities on 
the efficiency $\eta$ where the blue dots denote the simulated probabilities obtained with the sampling rejection technique. The red dots are outputs of the reconstruction algorithm; the solid line corresponds to the expression~\eqref{multimodeprob0a} with chosen values for the parameters $M$ and $\alpha$. 
In the right plots of each figure, the red line shows the reconstructed photon distribution while the black line shows the photon distribution given 
by~\eqref{multimodeprob0}. The fidelity $F$ of the reconstruction is also reported.\\
\indent
The figures show the ability of the no-click reconstruction algorithm to efficiently reproduce the occupation number state statistics not only for low average 
photon numbers, as already demonstrated in the range of visible light, but also for higher ones as required by the FEL scenario. This numerical evidence points to the feasibility of an experiment where, without incurring the difficulties inherent in photon counting with many photons, one may access the diagonal entries of the quantum state of FEL radiation in the energy representation.

\section{Discussion and Conclusion}
We have illustrated a novel approach to the study of the laser properties of the radiation emitted by an FEL  by accessing not the electromagnetic correlation functions, but rather the diagonal photon statistics of the quantum state of the radiation in the energy representation. 
The occupation numbers are expected to follow thermal statistics in the case of SASE-FELs and that of a displaced thermal distribution in the case of seeded-FELs.
The latter statistics are closer to that of laser light, with its the higher coherence with respect to its thermal features.\\
Due to the large average number of photons expected in concrete experimental set-ups, the occupation number statistics can be hardly accessed via photon counting. Therefore, we suggest the use of a Maximum Likelihood reconstruction algorithm based on no-click probabilities at multiple, variable detector efficiencies. 
The reconstruction power of the algorithm have been tested in numerical experiments and proved to extend well to the high photon numbers typical of FEL light.\\
\indent
The numerical evidence points to the feasibility of an experimental implementation of the strategy described in the previous sections. As regards this future possibility, here 
we only observe (some more details are provided in the Supplementary Material) that monolithically integrated, complementary metal-oxide-semiconductor (CMOS) single photon avalanche photodiodes (SPADs)  appear particularly well-suited to measure no-click probabilities due to their fast response and short dead-times, especially in the case of FELs with high repetition rates.
Furthermore, arrays of photon counting pixels can be easily integrated on chip allowing pixels to be selectively turned off thus directly controlling the detection efficiency.\\
\indent
The experimental confirmation of a thermal distribution of the occupation numbers of the energy states of the SASE-FEL radiation would confirm its classical features in agreement with the fact that its purported squeezing is an artefact due to photon counting drawbacks. On the other hand,  the experimental evidence of a Poissonian distribution of seeded-FEL radiation would have two main consequences. In the first place it would represent a first step towards the development of an $X$-ray quantum optics in connection with high-energy light-matter interactions. In the second place, it would spur the investigation of the very mechanism behind the seeded-FEL radiation, whether and how  the seeding laser imprints its coherence on the emitted light and whether a sub-Poissonian seeding might induce the appearance of  genuinely quantum features in it.

\section*{authorship statements}
\noindent
Conception and design of the project: F.B., S.O., R.F., F. P.\\
Analysis and interpretation of research data: G.P., S.O.\\
Drafting and critically revising the research output: F.B.,F.P.\\
Contributions to the analysis and interpretation: R.F., S.D.M, D.B., L.R.\\
F.B. and F.P. are corresponding authors for the theoretical, respectively experimental aspects of the manuscript.

\section*{acknowledgments}
F.P. would like to thanks William A. Barletta (MIT) for the many useful discussions about the FEL physics relevant to the present study.
F.B. and R.F. acknowledge that their research has been conducted within the framework of the Trieste Institute for Theoretical Quantum Technologies.
S.D.M. and G.P. acknowledge that the experimental spectra have been collected at the FERMI FEL facility, within the framework of the FERMI Team operation.

\newpage
\appendix
\onecolumngrid

\section{Quantum states of light}
\label{Sec:Coherence}

In this section we present quantum states of light that are relevant for describing the SASE and seeded-FEL radiation.

A monocromatic, linearly polarized plane-wave  of frequency $\omega$ is described by single-mode creation and annihilation operators $a$ and $a^\dag$ satisfying 
the commutation relations $[a\,,\,a^\dag]=1$.
The absence of photons of frequency $\omega$ is associated  with the vacuum state $\vert 0\rangle$ which is annihilated by the annihilation operator: $a\vert 0\rangle=0$,
whereas the so-called  number state, or Fock state, describing  $n$ photons with frequency $\omega$ is obtained by acting $n$ times on the vacuum with the creation operator $a^\dag$:
\begin{equation}
\label{E5}
\vert n\rangle =\frac{(a^\dag)^{n}}{\sqrt{n!}}\vert 0\rangle\ . 
\end{equation}
The number states are orthogonal $\langle n\vert m\rangle=\delta_{nm}$ and such that
\begin{equation}
\label{E6}
a\,\vert n\rangle\,=\,\sqrt{n}\,\vert n\rangle\ ,\qquad a^\dag\vert n\rangle\,=\,\sqrt{n+1}\,\vert n+1\rangle\ ,
\end{equation}
whence they possess definite photon number $N=a^\dag a$ and thus definite energy $E=\omega N$ ($\hbar=1$): 
\begin{equation}
\label{E7}
N\,\vert n\rangle\,=\,n\,\vert n\rangle\ .
\end{equation}
To the other extreme with respect to Fock states, one finds  the so-called coherent states $\vert\alpha\rangle$, with complex amplitudes $\alpha\in\mathbb{C}$; they are eigenstates of the annihilation operator and their  intensities are given by the their mean photon number: 
\begin{equation}
\label{E9}
a\,\vert\alpha\rangle=\alpha\,\vert\alpha\rangle\ ,\quad 
\vert\alpha\vert^2=\langle\alpha\vert a^\dag\,a\vert\alpha\rangle\ .
\end{equation}
Any quantum state of light described by mode-operators $a$ and $a^\dag$ is representable as a density matrix $\rho$ acting on the Fock space space spanned by the number states, which in that representation reads:
\begin{equation}
\label{rhonondiag}
\rho=\sum_{k,\ell}\rho_{k\ell}\, \vert k\rangle\langle \ell\vert\ ,\quad \rho_{k\ell}=\langle k\vert\rho\vert\ell]\rangle\ .
\end{equation}
In turn, any such density 
matrix  can always be written 
as~\cite{Glauber}:
\begin{equation}
\label{Pfunction}
\rho=\int_{\mathbb{C}}{\rm d}^2\alpha\,\mathcal{P}_\rho(\alpha)\,\vert\alpha\rangle\langle\alpha\vert\quad\hbox{with}\quad\int_{\mathbb{C}}{\rm d}^2\alpha\,\mathcal{P}_\rho(\alpha)=1\,
\end{equation}
where $\mathcal{P}_\rho(\alpha)$ is the so-called Glauber-Sudarshan P-function (see e.g.~\cite{Gerry} for details) associated with the quantum state $\rho$ and the integration is performed with respect to the real and imaginary parts of  $\alpha=\alpha_x+i\alpha_y$, ${\rm d}^2\alpha= {\rm d}\alpha_x\,{\rm d}\alpha_y$. 

Therefore any single-mode state of light reads as a linear combination of projectors $\vert\alpha\rangle\langle\alpha\vert$ onto coherent states with a function 
$\mathcal{P}_\rho(\alpha)$ that, though being normalized, need not necessarily amount to a smooth probability distribution over the complex plane.
Indeed, $\mathcal{P}_\rho(\alpha)$  may not only degenerate into a highly singular distribution, but even become non-positive.
As a consequence, a quantum state of light is said to be 
\begin{itemize}
\item
classical if the Glauber--Sudarshan P-function is a \textit{bona fide} distribution
\begin{equation}
\label{classicallight}
\mathcal{P}_\rho(\alpha)\geq 0\ ,
\end{equation}
with $\mathcal{P}_\rho(\alpha)$  a smooth function or no more singular than a Dirac delta.
\end{itemize}

For instance, as we shall see in Section~\ref{1coherentstate:sec}, for coherent states, that is when $\rho=\vert\beta\rangle \langle\beta\vert$, $\mathcal{P}_\rho(\alpha)$ reduces to a Dirac delta $\delta^2(\alpha)=\delta(\alpha_x)\delta(\alpha_y)$ on the complex plane, while, for thermal states $\mathcal{P}_\rho(\alpha)$  reduces to a Gaussian distribution. Instead, a quantum state of light is called 
\begin{itemize}
\item
non-classical or, in other words, genuinely quantum,  if the function 
$\mathcal{P}_\rho(\alpha)$ is either non-positive 
\begin{equation}
\label{quanyumlight}
\mathcal{P}_\rho(\alpha)\ngeq 0\ ,
\end{equation}
or more singular than a Dirac delta distribution. 
\end{itemize}

Indeed, the non-classical behaviour of Fock number states is accompanied 
by a highly singular distributional Glauber-Sudarshan $\mathcal{P}_\rho$ function, as shown in Appendix~\ref{app1:sec},
\begin{equation}
\label{P-funct2}
\mathcal{P}_n(\alpha)=\frac{\displaystyle{\rm e}^{\vert\alpha\vert^2}}{n!}\frac{\partial^{2n}}{\partial_\alpha^n\partial_{\alpha^*}^n}\delta^2(\alpha)\ .
\end{equation}
The P-function being an increasingly (with $n$) singular distribution, its behaviour has to be tested by  integration against suitably differentiable trial functions $g(\alpha,\alpha^*)$ vanishing at infinity, that is by computing
\begin{equation}
\label{P-funct3a}
\mathcal{P}_n[g]=\int_{\mathbb{C}^2}{\rm d}^2\alpha\,{\rm d}^2\alpha^*\, \mathcal{P}_n(\alpha)\, g(\alpha,\alpha^*)=\frac{\displaystyle{\rm e}^{\vert\alpha\vert^2}}{n!}\frac{\partial^{2n}g(\alpha,\alpha^*)}{\partial_\alpha^n\partial_{\alpha^*}^n}\ .
\end{equation} 
Evidently, these integrations can yield negative values even for positive trial functions, thus the P-function $\mathcal{P}_n(\alpha)$ cannot be a positive distribution, whence the Fock number states are highly non-classical.

We shall now examine some quantum states associated with different degrees of coherence and discuss swhich ones among them are good candidates for the quantum description of the radiation emitted by SASE and seeded-FELs.
Technical details on how to derive their mathematical properties are given in the Appendices.
 
\begin{description}
\item[Coherent states and laser radiation]
\label{1coherentstate:sec}

As we have seen in~\eqref{E9}, coherent states are eigenstates of the annihilation operator
and have thus the following expansion over the orthonormal basis of number states~\eqref{E5}:  
\begin{equation}
\label{cohst0}
\ket{\alpha}={\rm e}^{-\vert\alpha\vert^2/2}\,\sum_{n=0}^\infty\frac{\alpha^n}{\sqrt{n!}}\,\ket{n}\ ,\quad \alpha\in\mathbb{C}\ ,
\end{equation}
whence the associated number or energy distribution is Possonian:
\begin{equation}
\label{Poisson}
\vert\langle n\vert\alpha\rangle\vert^2={\rm e}^{-|\alpha|^2}\frac{|\alpha|^{2n}}{n!}\ .
\end{equation}

Though their quantum granularity is embodied by the Poissonian distribution~\eqref{Poisson}, coherent states $\rho_\gamma=\vert\gamma\rangle\langle\gamma\vert$ are nevertheless classical saccording to the definition comprising equation~\eqref{classicallight}. Indeed, (see Appendix~\ref{app1:sec}),  their Glauber-Sudarshan P-function $\mathcal{P}_\gamma(\alpha)$~\eqref{Pfunction} is a Dirac delta at the point $\gamma$ in the complex plane:
\begin{equation}
\label{P-funct6}
\mathcal{P}_\gamma(\alpha)=\delta^2(\alpha-\gamma)\ .
\end{equation}
Given the complex amplitude $\alpha=\alpha_x+i \alpha_y$,  its phase-angle $\varphi=\arctan \alpha_y/\alpha_x$ cannot be determined; therefore, the states of monocromatic laser light  are obtained by averaging uniformly over all possible $\varphi$.
Such a procedure yields a uniform mixture of coherent states, but also a Poissonian mixture of number state projectors~\cite{Enk}. 
Writing $\displaystyle\alpha_\varphi=\alpha\ {\rm e}^{i\varphi}$, $\alpha\geq 0$,
\begin{eqnarray}
\label{laserstate1}
\rho_L=\frac{1}{2\pi}\,\int_0^{2\pi}{\rm d}\varphi\,\ket{\alpha_\varphi}\bra{\alpha_\varphi}={\rm e}^{-\alpha^2}\,\sum_{m,n=0}^\infty\frac{\alpha^{m+n}}{\sqrt{m!n!}}\,\ket{n}\bra{m}\,
\frac{1}{2\pi}\,\int_0^{2\pi}{\rm d}\varphi\,{\rm e}^{i\varphi(n-m)}={\rm e}^{-\alpha^2}\,\sum_{n=0}^\infty\frac{\alpha^{2n}}{n!}\,\ket{n}\bra{n}.
\end{eqnarray}

\item[Single mode thermal light]
\label{1thermalstate:sec}

A single mode thermal state of light at temperature $T$ is described by the Gibbs density matrix (setting the Boltzmann constant $\kappa_B=1$)
\begin{equation}
\label{Gibbs}
\rho_{T}=\left(1-{\rm e}^{-\omega/T}\right)\, {\rm e}^{-\omega/T\, a^\dag\, a}=\frac{1}{1+n_{T}}\sum_{n=0}^\infty\left(\frac{n_T}{1+n_{T}}\right)^n\,
\vert n\rangle\langle n\vert\ ,\ n_T\equiv\langle N\rangle=\frac{1}{{\rm e}^{\omega/T}-1}\ .
\end{equation}
Then, one computes
\begin{equation}
\label{Gibbs2}
 \langle N^2\rangle=\frac{{\rm e}^{\omega/T}+1}{\Big({\rm e}^{\omega/T}-1\Big)^2}\ ,\qquad
\langle N^2\rangle-\langle N\rangle=2\,\langle N\rangle^2\ .
\end{equation} 
As regards the thermal Glauber-Sudarshan P-function, one gets a P-function which is a Gaussian distribution (see~\eqref{P-funct4aux} in Appendix~\ref{app1:sec}) 
so that the corresponding quantum state can thus be interpreted as a classical one:
\begin{equation}
\label{P-funct4}
\mathcal{P}_T(\alpha)=\frac{1}{\pi}\frac{\displaystyle {\rm e}^{-\frac{\vert\alpha\vert^2}{n_T}}}{n_T}\ .
\end{equation} 
Indeed, thermal light can be seen as a mixture of coherent states with respect to a Gaussian distribution in the amplitude modulus $\vert\alpha\vert=\sqrt{\alpha_x^2+\alpha_y^2}$ of the amplitude 
$\displaystyle\alpha=\vert\alpha\vert\,{\rm e}^{i\varphi}=\alpha_x+i\alpha_y$ and a uniform distribution with respect to the phase-angle $\varphi=\arctan\alpha_y/\alpha_x$:
\begin{equation}
\label{thst}
\rho_T=\int_{\mathbb{R}^2}{\rm d}\alpha_x{\rm d}\alpha_y\,
\frac{\displaystyle{\rm e}^{-(\alpha_x^2+\alpha_y^2)/n_T}}{\pi n_T}\,\vert\alpha_x+i\alpha_y\rangle\langle\alpha_x+i\alpha_y\vert\ .
\end{equation}

\item[Displaced thermal states]
\label{1dispthermalstate:sec}

The randomness of thermal states expressed by their Gaussian Glauber-Sudarshan P-function can be diminished by displacing them via the displacement operators $D(\alpha)$ in~\eqref{displop} of Appendix~\ref{app1:sec}:
\begin{equation}
\label{displth1}
\rho_{T,\alpha}:=D(\alpha)\,\rho_T\,D^\dag(\alpha)\ .
\end{equation}
Indeed, coherent states can also be obtained by displacing the vacuum state, $\vert\alpha\rangle=D(\alpha)\vert 0\rangle$; therefore, by means of $D(\alpha)$, thermal states acquire a coherent contribution; this is particularly evident for vanishing temperatures, when $n_T$ vanishes as well and thermal states behave as the vacuum state which is turned into a fully coherent state by $D(\alpha)$ as in~\eqref{displop3}. A more formal characterizaton of this interpretation 
follows by using~\eqref{thst} and~\eqref{DDD}; indeed, one gets
\begin{equation}
\label{thdispl}
\rho_{T,\alpha}=\int_{\mathbb{C}}{\rm d}^2\beta\frac{\displaystyle{\rm e}^{-\vert\beta\vert^2/n_T}}{\pi n_T}\,\vert\alpha+\beta\rangle\langle \alpha+\beta\vert=\int_{\mathbb{C}}{\rm d}^2\beta\frac{\displaystyle{\rm e}^{-\vert\beta-\alpha\vert^2/n_T}}{\pi n_T}\,\vert\beta\rangle\langle \beta\vert\ ,
\end{equation}
where one sees that the displacement introduces a coherent bias into the Gaussian distribution that originates chaotic thermal light.

A concrete consequence of this coherent bias emerges when one looks at the mean energy  which also gets a coherent contribution; indeed,
using~\eqref{displop}, one computes
\begin{equation}
\label{displthrmen}
\langle a^\dag a\rangle={\rm Tr}\left(\rho_{T,\alpha}\,a^\dag a\right)=n_T+|\alpha|^2\ .
\end{equation}
\end{description}

\section{No-click quantum state reconstruction}

We now illustrate how the structure of a photon state diagonal with respect to the number representation can be reconstructed  by means of  the empirically measured  probability of no-click events from photodetectors in on/off experiments with varying detector efficiencies $\eta$.

\subsection*{Monochromatic light}
\label{subsec1mode}

We first consider the case of a single mode radiation, a monochromatic beam of photons of frequency $\omega$ described by a density matrix as in~\eqref{rhonondiag}.
If photons are detected by $L$ detectors with different efficiencies  $\eta_1,\eta_2,\ldots,\eta_L$, one collects
$L$ linear equations 
\begin{equation}
P_\ell(\{\rho_n\}_n)=\sum_{k=0}^N(1-\eta_\ell)^k\,\rho_k\ ,\quad \ell=1,2,\ldots,L\ ,
\label{NoclickModel}
\end{equation}
that, as we shall presently show,  can be used to infer the $N$ unknown parameters $\rho_n$ via a Maximum Likelihood method.

\begin{rem}
\label{rem3.1}
The number $L$ of efficiencies to be considered has to be evaluated in reference with 
the specific reconstruction problem: in the case of visible light, as well as in the case of the simulations discussed in the main text, $L=20$ 
appears to be sufficient.
The fact that the number $N$ of unknowns is larger than the number of equations $L$ makes 
the problem underestimated with a non-unique solution, in general.
This issue can be overcome with an external energy constraint as discussed below. 
\end{rem}

For each efficiency $\eta_\ell$ one counts the 
no-click events $\mathcal{N}_\ell$ out of a total number $\mathcal{N}$  of impinging photons coming from a radiation beam described by a  state $\rho$, thus collecting $L$ empirical 
frequencies  
\begin{equation}
\label{freq}
f_\ell:=\frac{\mathcal{N}_\ell}{\mathcal{N}}\ .
\end{equation}
Usually, once the empirical frequencies $f_\ell$ of a set of independent events $x_\ell$ are measured, the event probabilities $p_\ell:=p(x_\ell)$ are derived from the log-likelihood function
\begin{equation}
\label{ML0}
L(\{p_n\}_n,\{f_n\}_n,\lambda)=\frac{1}{\mathcal{N}}\log \prod_\ell (p_\ell)^{\mathcal{N}_\ell}\,+\,\lambda\Big(\sum_\ell p_\ell-1\Big)\ ,
\end{equation} 
where the Lagrange multiplier $\lambda$ ensures normalization.
Then, maximizing with respect to the unknown $p_\ell$ one gets
\begin{equation}
\label{ML1}
\partial_{p_\ell}L(\{p_n\}_n,\{f_n\}_n,\lambda)=\frac{f_\ell}{p_\ell}\,+\,\lambda=0\  ,
\end{equation} 
whence, from $\sum_\ell f_\ell=1=\sum_\ell p_\ell$, one retrieves $\lambda=-1$ and $p_\ell=f_\ell$, as expected.

In the case of~\eqref{NoclickModel}, the unknown parameters to be found by optimization are the populations $\rho_n$ of the $n$-photon states, given the overall radiation state $\rho$. 
Then, one firstly introduces the normalized no-click probabilities
\begin{equation}
\label{prob}
\widetilde{P}_\ell(\{\rho_n\}_n):=\frac{P_\ell(\{\rho_n\}_n)}{P(\{\rho_n\}_n)}\ ,\quad \hbox{where}\quad P(\{\rho_n\}_n):=\sum_{\ell=1}^LP_\ell(\{\rho_n\}_n)\ ,
\end{equation}
and then the log-likelihood
\begin{equation}
\label{loglik}
L\left(\{\rho_n\}_n,\{f_n\}_n,\lambda\right):=\sum_{\ell=1}^L\,f_\ell\,\log \widetilde{P}_\ell(\{\rho_n\}_n)\,+\lambda\Big(\sum_{n=1}^N\rho_n-1\Big) \ .
\end{equation}
Setting $\partial_{\rho_n}L\left(\{\rho_n\}_n,\{f_n\}_n,\lambda\right)=0$, from $\sum_{k=0}^N\rho_k=1$, one obtains $\lambda=0$ and the following $N$ equalities
\begin{equation}
\label{fixedpoint1}
\frac{P(\{\rho_n\}_n)}{F}\, \sum_{\ell=1}^L\frac{(1-\eta_\ell)^n}{\sum_{k=1}^L(1-\eta_k)^n}\,\frac{f_\ell}{P_\ell(\{\rho_n\}_n)} =1\ ,
\qquad F:=\sum_{\ell=1}^L f_\ell\\ ,\quad n=1,2,\ldots, N\ .
\end{equation}
Setting
\begin{equation}
\label{fixedpoint1b}
\rho'_k=\frac{F}{P(\{\rho_n\}_n)}\,\rho_k\Longrightarrow P_\ell(\{\rho'_n\}_n)=\frac{F}{P(\{\rho_n\}_n)}\,P_\ell(\{\rho_n\}_n)
\end{equation}
one recasts~\eqref{fixedpoint1} as
\begin{equation}
\label{fixedpoint2}
\sum_{\ell=1}^L\frac{(1-\eta_\ell)^n}{\sum_{k=1}^L(1-\eta_k)^n}\,\frac{f_\ell}{P_\ell(\{\rho'_n\}_n)} =1\ ,
\end{equation}
where the quantities $\rho'_k$ are positive but not normalized:
\begin{equation}
\label{fixedpoint3}
\sum_{k=1}^N\rho'_k=\frac{F}{P(\{\rho_n\}_n)}\neq 1\ .
\end{equation}
Notice that the equality~\eqref{fixedpoint1} can be turned into a fixed point relation
\begin{equation}
\label{fixedpoint4}
\sum_{\ell=1}^L\frac{(1-\eta_\ell)^p}{\sum_{k=1}^L(1-\eta_k)^p}\,\frac{f_\ell}{P_\ell(\{\rho'_n\}_n)}\rho'_p\,=\,\rho'_p\ ,
\end{equation}
whence the probabilities 
\begin{equation}
\label{fixedpoint5}
\rho_p=\frac{\rho'_p}{\sum_{k=0}^N\rho'_k}
\end{equation}
can be achieved by the iteration procedure
\begin{equation}
\label{fixedpoint6}
\sum_{\ell=1}^L\frac{(1-\eta_\ell)^p}{\sum_{j=1}^L(1-\eta_j)^p}\,\frac{f_\ell}{P_\ell(\{\rho_n(h)\})_n}\,\frac{\rho_p(h)}{\sum_{k=1}^N\rho_k(h)}\,=\,\rho_p(h+1)\ ,
\end{equation}
where  $\displaystyle\frac{\rho_p(h)}{\sum_{k=0}^N\rho_k(h)}$ are the probabilities at the $h$-th iteration step. 

Due to the fact that in general the number of efficiencies $L$ is much smaller than the number $N$ of probabilities to be determined, the linear problem~\eqref{NoclickModel} is underestimated.  Especially in the multi-mode case to be treated in the next section, it could happen that more than one diagonal distribution $\{\rho_n\}_n$ may fit the same experimental no-click probabilities. 
In order to sort the true distribution out, it proves convenient to constrain the energy through one more Lagrange multiplier $\beta$, changing the functional $L\left(\{\rho_n\}_n,\{f_n\}_n,\lambda\right)$ in~\eqref{loglik} into 
\begin{equation}
\label{loglik-const}
L_\beta\left(\{\rho_n\}_n,\{f_n\}_n,\lambda\right):=L\left(\{\rho_n\}_n,\{f_n\}_n,\lambda\right)+\beta\Big(\sum_{k=1}^Nk\,\rho_k-E\Big) \ .
\end{equation}
The optimization procedure leads to the following constrained version of~\eqref{fixedpoint1}
\begin{equation}
\label{fixedpoint1-const}
\frac{P(\{\rho_n\}_n)}{F}\, \sum_{\ell=1}^L\frac{(1-\eta_\ell)^p}{\sum_{k=1}^L(1-\eta_k)^p-\frac{P(\{\rho_n\}_n)}{F}\beta(p-E)}\,\frac{f_\ell}{P_\ell(\{\rho_n\}_n)} =1\ .
\end{equation}
Then, the reconstruction algorithm proceeds as above, iteratively tuning $\beta$ starting from $\beta=0$ as suggested in~\cite{Brida2}, in order to eliminate configurations not compatible with the mean energy $E$.

\subsection*{Multi-mode case}
\label{multimodesec}

In the non-monocromatic case, let us suppose  the beam contains $M>1$ different independent modes corresponding to photons with $M$ different frequencies
$\omega_1,\omega_2,\ldots,\omega_M$. The quantum state of the beam is  then factorized:
\begin{equation}
\label{multimodestate}
\rho=\bigotimes_{j=1}^M\rho^{(j)}\ ,
\end{equation}
where $\rho^{(j)}$ denotes the state associated with the $j$-th mode.
Given a same detector efficiency $\eta$ for all modes, the joint probability of having $\bar n$ photons distributed over the $M$ modes together with no clicks reads
\begin{equation}
\label{multimodeprob0}
\Pi_{\bar n}(M,\eta)=\sum_{n_1+n_2+\cdots+n_M=\bar n}\left[\prod_{j=1}^M(1-\eta)^{n_j}\rho^{(j)}_{n_j}\right]\ ,
\end{equation}
where $n_j$, $j=1,2,\ldots,M$, is the occupation number of the mode of frequency $\omega_j$, $\rho^{(j)}_{n_j}$ is the probability of having $n_j$ photons with that frequency 
in the state $\rho^{(j)}$ and the summation is over all possible occupation numbers of those modes conditioned on the total number of photons being $\bar n$.
Then, the total probability of no-clicks is given by
\begin{equation}
\label{multimodeprob0a}
P_M(\eta)=\sum_{\bar n=0}^\infty \Pi_{\bar n}(M,\eta)\ ,
\end{equation}
Notice that normalization asks for
$$
\sum_{\bar n=0}^\infty\ \sum_{n_1+n_2+\cdots+n_M=\bar n}\ \left[\prod_{j=1}^M\rho^{(j)}_{n_j}\right]=1\ ,
$$ 
whence as in the one-mode case, one chooses a suitable accuracy parameter $\epsilon$ and truncates the summation at
$\bar n_{max}$ such that
\begin{equation}
\label{multimodeprob}
\sum_{\bar n=0}^{\bar n_{max}}\sum_{n_1+n_2+\cdots+n_M=\bar n}\prod_{j=1}^M\rho^{(j)}_{n_j}\geq 1-\epsilon\ .
\end{equation}
Then, varying over a number of quantum efficiencies $\eta_1,\eta_2\,\ldots,\eta_L$, $L\geq \bar n_{max}$,
one obtains the following multimode  system of equations similar to~\eqref{NoclickModel}:
\begin{equation}
P_M(\eta_j)=\sum_{\bar n=0}^{\bar n_{max}}\Pi_{\bar n}(M,\eta_j)\ ,\quad j=1,2,\ldots,L\ .
\label{NoclickModelM}
\end{equation}
The recursive algorithm can then reproduce the probabilities  $\Pi_{\bar n}(M,\eta_j)$ by means of the no-click probabilities $P_M(\eta_j)$ after renormalization as in~\eqref{prob}.

Notice that, by recursion, the probabilities $\Pi_{\bar n}(\eta)$ can be turned into discrete nested convolutions
\begin{eqnarray}
\label{multimodeprob1}
\Pi_{\bar n}(M,\eta)&=&\sum_{n_1=0}^{\bar n} (1-\eta)^{n_1}\rho^{(1)}_{n_1}\,\Pi_{\bar n-n_1}(M-1,\eta)=\sum_{n_1=0}^{\bar n}\sum_{n_2=0}^{\bar n-n_1}(1-\eta)^{n_1+n_2} \,\rho^{(1)}_{n_1}\rho^{(2)}_{n_2}
\, \Pi_{\bar n-n_1-n_2}(M-2,\eta)\\
\label{multimodeprob2}
&=&(1-\eta)^{\bar n}\sum_{n_1=0}^{\bar n}\sum_{n_2=0}^{\bar n-n_1}\cdots\sum_{n_{M-1}=0}^{\bar n-\sum_{k=1}^{M-2}n_k}\,\prod_{j=1}^{M} \rho^{(j)}_{n_j}\ ,\quad n_{M}
=\bar n-\sum_{j=1}^{M-1}n_j\ .
\end{eqnarray}

\subsubsection*{Multi-mode coherent states}
 
 As a concrete case, let us imagine that the photons are in coherent states of amplitudes $\alpha_j$ relative to the modes in the beam; 
 namely, from~\eqref{Poisson}, the occupation numbers read
\begin{equation}
\label{cohj}
\rho^{(j)}_{n_j}={\rm e}^{-\vert\alpha_j\vert^2}\,\frac{\vert\alpha_j\vert^{2n_j}}{n_j!}\ .
\end{equation}
In such a case, from~\eqref{multimodeprob2} and~\eqref{multimodeprob0a}
one obtains
\begin{eqnarray}
\label{multimodeprob3}
\Pi_{\bar n}(M,\eta)&=&(1-\eta)^{\bar n}{\rm e}^{-\sum_{j=1}^M\vert\alpha_j\vert^2}\,
\sum_{n_1=0}^N\sum_{n_2=0}^{N-n_1}\cdots\sum_{n_{M-1}=0}^{\bar n-\sum_{k=1}^{M-2}n_k}\,\prod_{j=1}^{M} \frac{\vert\alpha_j\vert^{2n_j}}{n_j!}\ ,\quad 
n_{M}=\bar n-\sum_{j=1}^{M-1}\,n_j\\
\label{multimodeprob4}
&=&{\rm e}^{-\sum_{j=1}^M\vert\alpha_j\vert^2}\,\frac{(1-\eta)^{\bar n}}{\bar n!}\,\Big(\sum_{k=1}^M\vert\alpha_j\vert^2\Big)^{\bar n}\ ,
\end{eqnarray}
namely a damped Poisson distribution, whence the no-click probabilities are the multimode Gaussians
\begin{equation}
\label{multimodeprob5}
P_M(\eta)=\sum_{\bar n=0}^{\bar n_{max}}\Pi_{\bar n}(M,\eta)\simeq {\rm e}^{-\eta\,\sum_{j=1}^M\vert\alpha_j\vert^2}\ .
\end{equation}
It is important to emphasize that the above result also holds when the coherent state projections $\rho^{(j)}$ are replaced by the uniform phase-averages 
$\rho_L$ in~\eqref{laserstate1}; indeed,
both these states have the same diagonal elements in  the Fock number state representation.  Therefore, $P_M(\eta)$ in~\eqref{multimodeprob5} represents the no-click probability for a multi-mode laser beam.

\subsubsection*{Multimode-displaced thermal states}
\label{displthalg:sec}

Let us suppose that the $M$-mode state consist of thermal states $\rho_{T,\alpha_j}$ at a same temperature $T$ but displaced by different amplitudes $\alpha_j$.
The single-mode occupation probability of the $n$-th Fock number state is (see Appendix~\ref{app2:sec})
\begin{equation}
\label{multimodeprob6}
\rho_{T,\alpha}(n)=\left(\frac{n_T}{1+n_T}\right)^n\,\frac{\displaystyle{\rm e}^{-\frac{\vert\alpha\vert^2}{1+n_T}}}{1+n_T}\,L_n\left(-\frac{\vert\alpha\vert^2}{n_T(1+n_T)}\right)\ ,
\end{equation}
where $L_n(x)=L^{(0)}_n(x)$ are the Laguerre polynomials.

Insert the displaced thermal states 
$\rho_{T,\alpha_j}$ in the place of all $\rho^{(j)}$ in~\eqref{multimodeprob1} and consider the last sum over $n_{M-1}$:
$$
\Sigma_M:=\sum_{n_{M-1}=0}^{\bar n-\sum_{k=1}^{M-2}n_k}\,\rho^{(M-1)}_{n_{M-1}}\rho^{(M)}_{\bar n-\sum_{k=1}^{M-1}n_k}\ .
$$
Using~\eqref{multimodeprob6}, one obtains
\begin{eqnarray*}
\Sigma_M&=&\frac{1}{(1+n_T)^2}\,{\rm e}^{-2\frac{\vert\alpha\vert^2}{1+n_T}}\, \left(\frac{n_T}{1+n_T}\right)^{\bar n-\sum_{k=1}^{M-2}n_k}\,\times\\
&\times&\sum_{n_{M-1}=0}^{\bar n-\sum_{k=1}^{M-2}n_k}\,L_{n_{M-1}}\left(-\frac{\vert\alpha_{M-1}\vert^2}{n_T(1+n_T)}\right)\,L_{\bar n-\sum_{k=1}^{M-2}n_k\,-\,n_{M-1}}\left(-\frac{\vert\alpha_M\vert^2}{n_T(1+n_T)}\right)\ .
\end{eqnarray*}
Using the sum rule~\cite{Gradshtein}
\begin{equation}
\label{sumrule}
L_n^{(\alpha+\beta+1)}(x+y)=\sum_{k=0}^nL^{(\alpha)}_k(x)\,L^{(\beta)}_{n-k}(y)\ ,
\end{equation}
one finally gets
\begin{equation}
\label{multimodeprob7}
\Sigma_M=\frac{1}{(1+n_T)^2}\,{\rm e}^{-2\frac{\vert\alpha\vert^2}{1+n_T}}\, \left(\frac{n_T}{1+n_T}\right)^{\bar n-\sum_{k=1}^{M-2}n_k}\,
L^{(1)}_{\bar n-\sum_{k=1}^{M-2}n_k\,-\,n_{M-1}}\left(-\,\frac{\vert\alpha_{M-1}\vert^2+\vert\alpha_M\vert^2}{n_T(1+n_T)}\right)\ .
\end{equation}
Iterating this procedure yields
\begin{equation}
\label{multimodeprob8}
\Pi_{\bar n}(M,\eta)=\frac{1}{(1+n_T)^M}\,{\rm e}^{-\frac{\sum_{j=1}^M\,\vert\alpha_j\vert^2}{1+n_T}}\, \left(\frac{(1-\eta)n_T}{1+n_T}\right)^{\bar n}\,
L^{(M-1)}_{\bar n}\left(-\frac{\sum_{j=1}^M\vert\alpha_j\vert^2}{n_T(1+n_T)}\right)\ .
\end{equation}
Thence the no-click probability at efficiency $\eta$ results
\begin{equation}
\label{multimodeprob9}
P_M(\eta)\simeq\sum_{\bar n=0}^\infty P^{(M)}_{\bar n}(\eta)=\frac{1}{(1+\eta n_T)^M}\,
\exp\Big(-\frac{\eta\,\sum_{j=1}^M\,\vert\alpha_j\vert^2}{1+\eta\, n_T}\Big)\ .
\end{equation}
The last equality follows from using the associated Laguerre polynomials generating equation
\begin{equation}
\label{genfunct}
(1-t)^{-1-\lambda}\,\exp\Big(-\frac{x\,t}{1-t}\Big)=\sum_{n=0}^\infty L^{(\lambda)}_n(x)\, t^n\ .
\end{equation}

\section{Detectors}

When considering the detectors that should measure no-click probabilities, several requirements need be satisfied. The detector should have fast response ($< 1\,ns$) and a relatively small dead time (ideally $<100\, ns$) to accommodate for the laser dynamics, especially in the case of FELs with high repetition rates. The quantum efficiency should be high ($>10$).
One of the most employed single photon detectors are photomultiplier tubes \cite{Howe2006}. The figures of merit for photomultipliers are close to those specified above, even if response times are often larger than $1\, ns$ and dead times in the range between $100\, ns$ and $1\, \mu s$. One disadvantage of photomultiplier tubes is the need to bias the devices at voltages of the order of $1\, kV$, making gated operation extremely difficult.  

Recent years have seen the development of monolithically integrated CMOS detectors. Monolithic integration of the photosensitive area with the readout electronics offers unprecedented accuracy in the processing of the detected signal and engineering freedom on the sensor characteristics in term of size and electronic response. Fully depleted p-n junctions have been used for X-ray detection and have sensitivities that can reach the single photon level \cite{Giampaolo2019,Hirono2016}. Different materials can be employed for the sensor area via wafer bonding \cite{Ruat2012}.

Monolithic CMOS detectors can operate in Geiger mode in the case of CMOS single photon avalanche photodiodes (SPADs) \cite{ghioni17}. CMOS SPADS have very fast response times (generally $<100\, ps$) with short dead times (of the order of some tens of $ns$) and can be operated at low voltage of the order of $10\, V$. 
Furthermore their integration can ensure that arrays of photon counting pixels can be easily integrated on chip to the side of the electronic circuits used for counting \cite{braga14}.  Such pixels can be as small as $10\, \mu m$ in size, giving the possibility of engineering arrays comprising tens or hundreds of single photon counting pixels within the FEL spot diameter. The counts from such an array can then be summed by the on-chip electronics (in a configuration similar to silicon photomultipliers \cite{Conca2020}) 
and pixels can be selectively turned off thus directly controlling the detection efficiency.

\section{P-function: coherent, thermal and number states}
\label{app1:sec}

Coherent states form a continuous non-orthogonal family, 
\begin{equation}
\label{cohscpr}
\langle\alpha\vert\beta\rangle={\rm e}^{\displaystyle-\frac{1}{2}(\vert\alpha\vert^2+\vert\beta\vert^2-2\alpha^*\beta)}\ ,
\end{equation}
which is however over-complete  in the sense that the corresponding projectors integrate to the identity operator
\begin{equation}
\label{hypercompl}
\int_{\mathbb{C}}\frac{{\rm d}^2\alpha}{\pi}\,\ket{\alpha}\bra{\alpha}=\mathbb{I}\ .
\end{equation}
Coherent states are obtained by displacing the vacuum $\vert 0\rangle$, 
\begin{equation}
\label{displop3}
D(\alpha)\ket{0}=\ket{\alpha}\ ,
\end{equation}
by means of the displacement operators
\begin{equation}
\label{displop}
D(\alpha)={\rm e}^{\alpha\,a^\dag\,-\,\alpha^*\,a}={\rm e}^{-\vert\alpha\vert^2/2}\,{\rm e}^{\alpha\,a^\dag}\,{\rm e}^{-\alpha^*\,a}\ ,
\end{equation} 
which are such that
\begin{equation}
\label{DDD}
D(\beta)\,D(\alpha)={\rm e}^{(\beta^*\alpha-\beta\alpha^*)/2}\,D(\alpha+\beta)\ , \quad
D^\dag(\beta)\,D(\alpha)\,D(\beta)={\rm e}^{\beta^*\alpha-\beta\alpha^*}\,D(\alpha)\ ,\quad D^\dag(\alpha)\,a\,D(\alpha)=a+\alpha\ .
\end{equation}

Let $\rho$ be the density matrix of a monocromatic light beam; its  Glauber-Sudarshan representation reads as
a continuous linear combination of coherent-state projectors
$$
\rho=\int_{\mathbb{C}}{\rm d}^2\alpha\, \mathcal{P}_\rho(\alpha)\,\vert\alpha\rangle\langle\alpha\vert\ .
$$
Using the scalar product~\eqref{cohscpr}, with $\alpha=x+iy$ and $\beta=u+iv$, one obtains
\begin{eqnarray}
\nonumber
{\rm e}^{\vert\beta\vert^2}\,\langle-\beta\vert\rho\vert\beta\rangle&=&\int_{\mathbb{C}}{\rm d}^2\alpha\, \mathcal{P}_\rho(\alpha)\, 
{\rm e}^{-\vert\alpha\vert^2-\alpha]\beta^*+\alpha^*\beta}\\
\label{app1.1}
&=&\int_{\mathbb{R}^2}{\rm d}x{\rm d}y\,\mathcal{P}_\rho(x+iy)\, {\rm e}^{-x^2-y^2}\,{\rm e}^{2i(xv-yu)}\ ,
\end{eqnarray}
whence, by anti-Fourier transformation,
\begin{equation}
\label{app1.2}
\int_{\mathbb{R}^2}\frac{{\rm d}(2u){\rm d}(2v)}{4\pi^2}\,{\rm e}^{u^2+v^2}\,\langle-u-iv\vert\rho\vert u+iv\rangle\,{\rm e}^{2i(xv-yu)}
={\rm e}^{-x^2-y^2}\,\mathcal{P}_\rho(x+iy)\ .
\end{equation}
Then, the Glauber-Sudarshan P-function  $\mathcal{P}_\rho(\alpha)$ can be expressed by means of the matrix elements  $\langle-\beta\vert\rho\vert\beta\rangle$: 
\begin{equation}
\label{P-funct1}
\mathcal{P}_\rho(\alpha)=\frac{\displaystyle{\rm e}^{\vert\alpha\vert^2}}{\pi^2}\int_{\mathbb{C}}{\rm d}^2\beta\, {\rm e}^{|\beta|^2}\,{\rm e}^{\alpha\beta^*-\alpha^*\beta}\,\langle-\beta\vert\rho\vert\beta\rangle\ ,
\end{equation}
where $\vert\pm\beta\rangle$ are coherent states. 

If $\rho$ is itself a coherent state $\rho_\gamma=\vert\gamma\rangle\langle\gamma\vert$, the scalar product~\eqref{cohscpr} yields
\begin{equation}
\label{P-funct5}
\langle-\beta\vert\rho_\gamma\vert\beta\rangle={\rm e}^{-\vert\beta\vert^2-\vert\gamma\vert^2-\gamma\beta^*+\gamma^*\beta}\ .
\end{equation}
Using~\eqref{P-funct1} and the fact that, with $\alpha=x+iy$ and $\beta=u+iv$,
\begin{equation}
\label{P-funct3}
\frac{1}{\pi^2}\int_{\mathbb{C}}{\rm d}^2\beta\,{\rm e}^{\alpha\beta^*-\alpha^*\beta}=\int_{-\infty}^{+\infty}\int_{-\infty}^{+\infty}\frac{{\rm d}x}{\pi}\frac{{\rm d}y}{\pi} {\rm e}^{2i(yu-xv)}=
\delta(x)\delta(y)=\delta^2(\alpha)\ ,
\end{equation}
one then obtains
\begin{equation}
\label{P-funct6}
\mathcal{P}_{\rho_\gamma}(\alpha)=\frac{\displaystyle{\rm e}^{\vert\alpha\vert^2-\vert\gamma\vert^2}}{\pi^2}\int_{\mathbb{C}^2}{\rm d}^2\beta\ {\rm e}^{-(\gamma-\alpha)\beta^*+(\gamma^*-\alpha^*)\beta}=\delta^2(\alpha-\gamma)\ .
\end{equation}
For thermal states as in~\eqref{Gibbs}, by Gaussian integration one obtains
\begin{equation}
\label{P-funct4aux}
\langle-\beta\vert\rho_T\vert\beta\rangle=\frac{\displaystyle{\rm e}^{-\vert\beta\vert^2\frac{1+2n_T}{1+n_T}}}{1+n_T}\Longrightarrow
\mathcal{P}_T(\alpha)=\frac{1}{\pi}\frac{\displaystyle {\rm e}^{-\frac{\vert\alpha\vert^2}{n_T}}}{n_T}\ .
\end{equation}
On the other hand, when $\rho=\vert n\rangle\langle n\vert$, one computes
\begin{equation}
\label{P-funct2}
\langle-\beta\vert\rho\vert\beta\rangle=\frac{\displaystyle{\rm e}^{-|\beta|^2}}{\pi^2}(-)^n\frac{\vert\beta\vert^2}{n!}\ ,
\end{equation}
whence, using~\eqref{P-funct3}, the P-function becomes a highly singular distribution:
\begin{equation}
\label{P-funct2}
\mathcal{P}_n(\alpha)=\frac{\displaystyle{\rm e}^{\vert\alpha\vert^2}}{\pi^2}\int_{\mathbb{C}}{\rm d}^2\beta\, (-)^n\frac{\vert\beta\vert^2}{n!}\,{\rm e}^{\alpha\beta^*-\alpha^*\beta}
=\frac{\displaystyle{\rm e}^{\vert\alpha\vert^2}}{\pi^2\,n!}\frac{\partial^{2n}}{\partial_\alpha^n\partial_{\alpha^*}^n}
\int_{\mathbb{C}}{\rm d}^2\beta\,{\rm e}^{\alpha\beta^*-\alpha^*\beta}=\frac{\displaystyle{\rm e}^{\vert\alpha\vert^2}}{n!}\frac{\partial^{2n}}{\partial_\alpha^n\partial_{\alpha^*}^n}\delta^2(\alpha)\ .
\end{equation}

\section{Displaced thermal states representation}
\label{app2:sec}

Sometimes, instead of their Glauber-Sudarshan representation in terms of projectors onto coherent states,  it proves convenient to represent quantum states by means of their characteristic function ${\rm Tr}( \rho\,D(-\alpha))$ and of the  displacement operators. 
In the case of thermal states, this representation reads
\begin{equation}
\label{displth2}
\rho_T=\int_{\mathbb{C}}{\rm d}^2\beta\,\frac{1}{\pi}{\rm e}^{-\vert\beta\vert^2\,(\frac{1}{2}+n_T)}\ D(\beta)\ ,
\end{equation}
whence, using the algebraic relation~\eqref{DDD}, it yields
\begin{equation}
\label{displth2}
\rho_{T,\alpha}=\int_{\mathbb{C}}{\rm d}^2\beta\,\frac{1}{\pi}{\rm e}^{-\vert\beta\vert^2\,(\frac{1}{2}+n_T)}\ {\rm e}^{\beta\alpha^*-\beta^*\alpha}D(\beta)
\end{equation}
for displaced thermal states.
These representations are obtained by firstly noticing that any single-mode density matrix $\rho$ can be written as
\begin{equation}
\label{app2.1}
\rho=\int_{\mathbb{C}}\frac{{\rm d}^2\beta}{\pi}\, {\rm Tr}\left(\rho\,D^\dag(\alpha)\right)\, D(\alpha)\ .
\end{equation}
This follows from using 
$$
D^\dag(\alpha)\ D(\beta)={\rm e}^{(\alpha^*\beta-\alpha\beta^*)/2}\ D(\beta-\alpha)\ ,
$$
and the overcompleteness relation~\eqref{hypercompl} to prove that
$$
{\rm Tr}\left(D^\dag(\beta)\,D(\alpha)\right)=\int_{\mathbb{C}}\frac{{\rm d}^2\gamma}{\pi}\,\langle\gamma\vert D^\dag(\beta)\,
D(\alpha)\vert\gamma\rangle=\pi\,\delta^2(\alpha-\beta)\ .
$$
Then,  using the P-function~\eqref{P-funct4} and the Glauber-Sudarshan representation together with the coherent state overcompleteness and Gaussian integration,  yield
\begin{equation}
\label{app2.2}
{\rm Tr}\left(D^\dag(\beta)\,\rho_T\right)=\frac{1}{\pi\,n_T}\int_{\mathbb{C}}{\rm d}^2\gamma\ {\rm e}^{-\vert\gamma\vert^2/n_T}\,\langle\gamma\vert D^\dag(\beta)\vert\gamma\rangle=\exp\left(-\vert\beta\vert^2\Big(\frac{1}{2}+n_T\Big)\right)\ .
\end{equation}


Given a displacement operator $D(\alpha)$ and a coherent state $\vert\beta\rangle=D(\beta)\vert 0\rangle$ (see~\eqref{displop}), 
using~\eqref{hypercompl} and then~\eqref{cohst0}, one computes in two ways the scalar products with Fock number states $\ket{n}$, $a^\dag a\ket{n}=n\ket{n}$:
\begin{eqnarray*}
\langle n\vert D(\alpha)\vert\beta\rangle&=&{\rm e}^{(\alpha\beta^*-\alpha^*\beta)/2}\,\langle n\vert D(\alpha+\beta)\vert0\rangle=
\frac{(\alpha+\beta)^n}{\sqrt{n!}}\,{\rm e}^{-(\vert\alpha\vert^2+\vert\beta\vert^2+2\alpha^*\beta)/2}\\
&=&{\rm e}^{-\vert\beta\vert^2/2}\,\sum_{m=0}^\infty\frac{\beta^m}{\sqrt{m!}}\, \langle n\vert D(]\alpha)\vert m\rangle\ .
\end{eqnarray*}
It thus follows that 
$$
(\alpha+\beta)^n\,{\rm e}^{-\alpha^*\beta}={\rm e}^{\vert\alpha\vert^2/2}\sum_{m=0}^\infty\sqrt{\frac{n!}{m!}}\beta^m\,\langle n\vert\, D(\alpha)\,\vert m\rangle\ .
$$
By setting $\beta=\alpha\,y$, the equality reads
$$
(1+y)^n\,{\rm e}^{-y\vert\alpha\vert^2}=\sum_{m=0}^\infty\left({\rm e}^{\vert\alpha\vert^2/2}\,\alpha^{m-n}\,\sqrt{\frac{n!}{m!}}\,\langle n\vert\, D(\alpha)\,\vert m\rangle\right)
\,y^m\ ,
$$
whose right hand side is recognizable as one of the generating functions of the associated 
Laguerre polynomials $L^{(\alpha)}_n(y)$~\cite{Gradshtein},
$$
(1+t)^\lambda\,{\rm e}^{-xt}=\sum_{m=0}^\infty L^{(\lambda-m)}_m(x)\, t^m\ .
$$
The identification 
\begin{equation}
\label{app3.1}
{\rm e}^{\vert\alpha\vert^2/2}\alpha^{m-n}\sqrt{\frac{n!}{m!}}\,\langle n\vert D(\alpha)\vert m\rangle=L^{(n-m)}_m(\vert\alpha\vert^2)
\end{equation}
finally yields
\begin{equation}
\label{app3.2}
\langle n\vert D(\alpha)\vert m\rangle={\rm e}^{-\vert\alpha\vert^2/2}\alpha^{n-m}\sqrt{\frac{m!}{n!}}\,L^{(n-m)}_m(\vert\alpha\vert^2)\ .
\end{equation}


From~\eqref{displth2} and~\eqref{app3.2} one gets
\begin{eqnarray}
\nonumber
\rho_{T,\alpha}(n)&=&\langle n\vert \rho_{T,\alpha}\vert n\rangle=\int_{\mathbb{C}}{\rm d}^2\beta\,\frac{1}{\pi}{\rm e}^{-\vert\beta\vert^2\,(\frac{1}{2}+n_T)}\ {\rm e}^{\beta\alpha^*-\beta^*\alpha}\langle n\vert D(\beta)\vert n\rangle\\
\label{app4.1}
&=&\int_{\mathbb{C}}{\rm d}^2\beta\,\frac{1}{\pi}{\rm e}^{-\vert\beta\vert^2\,(\frac{1}{2}+n_T)}\ {\rm e}^{\beta\alpha^*-\beta^*\alpha}{\rm e}^{-\vert\alpha\vert^2/2}\,L_n(\vert\alpha\vert^2)\ ,
\end{eqnarray}
where $L_n(x)=L^{(0)}_n(x)$ are the Laguerre polynomials.

By passing to polar coordinates, ${\rm d}^2\beta=r{\rm d}r{\rm d}\varphi$ and setting $\displaystyle\alpha=\vert\alpha\vert {\rm e}^{i\psi}$, one rewrites
 \begin{equation}
\label{app4.2}
\rho_{T,\alpha}(n)=\int_0^\infty \frac{{\rm d}}{\pi}\,r\,{\rm e}^{-r^2(1+n_T)}\,L_n(r^2)\,\int_0^{2\pi} {\rm d}\varphi\, {\rm e}^{2ir\vert\alpha\vert\sin(\varphi-\psi))}\ .
\end{equation}
The last integral equals the Bessel function $J_0(2r\vert\alpha\vert)$ so that one can use the relation~\cite{Gradshtein}
$$
\int_0^{+\infty}{\rm d}x\, x\, {\rm e}^{-\alpha x^2/2} \,L_n(\beta x^2/2)\, J_0(xy)=\frac{(\alpha-\beta)^n}{x^{n+1}}\,{\rm e}^{y^2/(2\alpha)}\, L_n\left(\frac{\beta y^2}{2\alpha(\beta-\alpha)}\right)
$$
and get 
$$
\rho_{T,\alpha}(n)=\left(\frac{n_T}{1+n_T}\right)^n\,\frac{\displaystyle{\rm e}^{-\frac{\vert\alpha\vert^2}{1+n_T}}}{1+n_T}\,L_n\left(-\frac{\vert\alpha\vert^2}{n_T(1+n_T)}\right)\ .
$$

\end{document}